# Is A 15-minute City within Reach in the United States? An Investigation of Activity-Based Mobility Flows in the 12 Most Populous US Cities

Tanhua Jin[a,b], Kailai Wang[c*], Yanan Xin[b*], Jian Shi[d], Ye Hong[b], Frank Witlox[a, e]

[a] *Department of Geography, Ghent University, Belgium*

[b] *Institute of Cartography and Geoinformation, ETH Zurich, Switzerland*

[c] *Department of Construction Management, University of Houston, USA*

[d] *Department of Engineering Technology, University of Houston, USA*

[e] *Department of Geography, University of Tartu, Estonia*

**Corresponding authors:** [c*] kwang43@central.uh.edu, 4230 Martin Luther King Boulevard. #300, Houston, TX 77204, USA; [b*] yanxin@ethz.ch, Stefano-Franscini-Platz 5, CH-8093 Zurich, Switzerland

**Abstract**

Enhanced efforts in the transportation sector should be implemented to mitigate the adverse effects of $CO_2$ emissions resulting from zoning-based planning paradigms. The innovative concept of the 15-minute city, with a focus on proximity-based planning, holds promise in minimizing unnecessary travel and advancing the progress toward achieving carbon neutrality. However, an important research question that remains insufficiently explored is: to what extent is a 15-minute city concept within reach for US cities? This paper establishes a comprehensive framework to evaluate the 15-minute city concept using SafeGraph Point of Interest (POI) check-in data in the 12 most populous US cities. The results reveal that residents are more likely to rely on cars due to the fact that most of their essential activities are located beyond convenient walking, cycling, and public transit distances. However, there is significant potential for the implementation of the 15-minute city concept, as most residents' current activities can be accommodated within a 15-minute radius by the aforementioned low-emission modes of transportation. When comparing cities, it appears that achieving a 15-minute walking city is more feasible for metropolises like New York City, San Francisco, Boston, and Chicago, while proving to be challenging for cities such as Atlanta, Dallas, Houston, and Phoenix. In inter-group comparisons, neighborhoods with a high proportion of White residents and a high median income tend to have more accessible POIs, with a high percentage of activities conducted within a 15-minute radius. White residents also have greater potential to satisfy their daily activities by walking/cycling/public transit trips within a 15-minute travel time, thus resulting in greater potential in $CO_2$ reduction compared to African Americans. Our findings can offer policymakers insight into how far US cities are away from the 15-minute city and the potential $CO_2$ emission reduction they can expect if the concept is successfully implemented.

## 1. Introduction

The report on the assessment of climate change released by the Intergovernmental Panel on Climate Change (IPCC) in 2021 emphasized that global temperatures are projected to rise by 2.7 °C by 2050. The projected temperature increase is substantially higher than the preferred goal of a maximum 1.5 °C increase, which was stated in the Paris Agreement (Intergovernmental Panel on Climate Change, 2021). Since 14% of annual global emissions are attributed to the transportation sector, of which 72% are from road vehicles (United Nations, 2021; Wu et al., 2023), enhanced efforts and immediate climate actions in this sector become imperative to mitigate the adverse effects of climate change (Hurlimann et al., 2021).

Existing literature has shown that the zoning-based urban planning paradigms which are predominant throughout the past century often resulted in the segregation of residential areas from commercial, retail, industrial, and entertainment zones (Raman & Roy, 2019). This spatial division led to increased travel distances for accessing daily necessities, resulting in reliance on automobiles and subsequently higher levels of $CO_2$ emissions. To address the drawbacks of zoning-based urban planning in terms of energy consumption and $CO_2$ emissions, the *15-minute city* concept has gained popularity (Abdelfattah et al., 2022; Büttner et al., 2022; Moreno et al., 2021; Pozoukidou et al., 2021). This visionary concept centers on proximity-based planning, with the primary goal of minimizing unnecessary travel and commute distances, while concurrently promoting a shift away from private vehicle usage in favor of more sustainable transportation alternatives (Di Marino et al., 2023; Silva et al., 2023). The key idea of the 15-minute city is to structure urban areas in a way that essential services are conveniently located within a 15-minute walking or cycling distance for residents (Wu et al., 2021).

The concept of a 15-minute city was first introduced in 2016 by Carlos Moreno as a framework to combat greenhouse gas (GHG) emissions (Moreno, 2016). Its main objective is to establish self-sufficient neighborhoods with the essential functions of living, working, commerce, healthcare, education, and entertainment by decentralizing urban operations and services (Moreno, 2016). The primary objective of the 15-minute city concept is to establish self-sustaining neighborhoods where essential functions such as living, working, commerce, healthcare, education, and entertainment are all within reach. This can be achieved through the decentralization of urban operations and services (Moreno, 2016). These socially connected and functionally mixed neighborhoods are designed at a human scale, ensuring that residents have adequate access to daily necessities by walking and cycling, reducing their reliance on private vehicles and consequently lowering carbon emissions (Khavarian-Garmsir et al., 2023). The COVID-19 pandemic has further amplified the appeal of the 15-minute city concept (Allam et al., 2022). It underscores the importance of creating safe and enjoyable pedestrian and cycling environments and emphasizes the effective implementation of high-quality and affordable public transportation systems as urban resilience against pandemics. These measures are essential in mitigating the adverse impacts of potential future epidemics (Gaglione et al., 2022). In the context of both climate change and the pandemic, numerous megacities across the world have attempted to embrace the 15-minute city concept. Paris is one of the first cities that has implemented an *"x-minute"* city strategy, and similar initiatives have also been successfully adopted in various regions, such as superblocks in Barcelona, high streets in the United Kingdom, and a 20-minute town in Shanghai (Logan et al., 2022). Nevertheless, implementing the 15-minute city concept in US cities faces considerable challenges, primarily due to the fact that Americans generally travel an average distance of 7 to 9 miles for daily activities (Lu & Diab, 2023). According to the 2017 US National Household Travel

Survey, the average distances of shopping, personal business, and recreation trips are 7.3 miles, 6.8 miles, and 8.7 miles, respectively. These distances significantly exceed what one could reasonably cover on foot or by bicycle within the concept of a 15-minute city. To address the possible problems and promote more sustainable transportation practices, Portland, Oregon, and Detroit, Michigan, have advocated for a maximum of 20 minutes to walk, bike, or use public transportation to access all amenities, in 2010 and 2016, respectively (City of Portland, 2010; Twenty-Minute Neighborhoods, 2016). Similarly, Tempe, Arizona, promotes a strategy of 'one-mile walking distance, a four-mile bicycle ride, or a 20-minute transit trip' (Arizona State University, 2021), while Seattle, Washington, endorses a range of distance and time measures contingent on zoning (Seattle City Council, 2020).

Given the strong reliance on cars in the US, it might be more pragmatic to expand the 15-minute city concept to encompass private vehicles or public transit as an intermediate step toward transitioning into active-mode-driven 15-minute cities. For instance, both Portland and Detroit in the USA have already extended the 15-minute city concept to include public transportation (City of Portland, 2010; Twenty-Minute Neighborhoods, 2016). Despite the growing popularity of the 15-minute city concept, there remains a paucity of research dedicated to evaluating its feasibility while considering the actual activity demands of residents. To this end, most previous studies have concentrated on quantifying the number of activity locations or services (i.e., Points of Interest (POIs)) accessible within 15-minute buffer zones (Di et al., 2023; Ferrer-Ortiz et al., 2022), leaving the following research questions largely unexplored:

1. How many activities are conducted by residents within 15-minute travel time?
2. Based on the people's realistic activity demand and current supplies of activity locations, how many current activities can be satisfied within 15-minute travel time? For those activity locations that are not within the 15-minute area, to what extent can residents find alternative activity locations within the 15-minute area?

3. Moreover, if residents conduct their activities that are not within 15-minute area by visiting those alternative activity locations within the 15-minute area, how many trip distances and $CO_2$ emissions can be reduced?

This paper presents a comprehensive framework for empirically evaluating the concept of the 15-minute city using large-scale POI check-in data in the 12 most populous US cities. The contributions can be summarized as follows. First, unlike previous studies that solely focused on accessible activity locations or services (i.e., POIs) within 15 minutes, this research extends further and assesses both the supplies and demands of activities within 15 minutes. Additionally, it examines the potential to implement a 15-minute city and reduce trip distances and $CO_2$ emissions. Second, in contrast to earlier research that mainly concentrated on walking and cycling within the 15-minute city context, this study considers multiple travel modes, including walking, cycling, public transit, and even car use. By embracing various transportation options, the extended concept of a 15-minute city emphasizes the importance of promoting localized neighborhoods while reducing trip distances and $CO_2$ emissions.

The rest of this paper is organized as follows. Section 2 describes the data sources and indicators to portray a 15-minute city. Section 3 discusses the overall assessment of the 15-minute city concept in US cities. Section 4 analyzes the inter-city and inter-group comparisons on the assessment of the 15-minute city. Section 5 examines to what extent the 15-minute city concept can lead to a reduction of trip distances and $CO_2$ emissions. Section 6 summarizes this research and proposes future research avenues.

## 2. Data and Methodology

*2.1 Data processing*

This research utilized POI check-in data obtained from SafeGraph[1], a data company specializing in the aggregation of anonymized location data from smartphone applications. The data was collected between June and October 2022, encompassing approximately 19 million smartphone devices across all 50 states and the District of Columbia. SafeGraph's dataset enabled the tracking of individuals' movements between their residences and Points of Interest (POIs). For privacy protection, residential locations were aggregated at the level of Census Block Groups (CBGs), which are relatively small administrative units containing a population range of 600 to 3,000 people. This dataset contains the connections between CBGs and activity locations, with the volume of visits for each activity location recorded. To ensure data quality and reliability, SafeGraph excluded home CBGs with less than five recorded devices at the POI during the entire month. Moreover, to avoid double counting, parent POIs, which encompass a small fraction of POIs in the dataset and include visits from their associated 'child' POIs, were also excluded. For instance, a parent POI could be a shopping mall that includes numerous small stores. SafeGraph data is geographically representative, covering 99.6% of CBGs (n = 210,288) in the US. Previous research has demonstrated that the dataset exhibits no systematic overrepresentation of individuals from CBGs across different regions or with distinct socio-demographic compositions, such as races, income levels, and education levels (Coleman et al., 2022; Coston et al., 2021).

The concept of the 15-minute city encompasses various essential functions in people's daily lives, including groceries, education, greenspaces, restaurants, religious establishments, healthcare facilities, recreational areas, and various services. Consequently, this study has extracted corresponding POI check-in data related to these essential activities from the original dataset. The dataset comprises a total of over 6.43 million POIs, with 35.8% of them categorized as essential POIs. Among these essential POIs, restaurants (28.9%) make up the largest portion, followed by service establishments (17.1%), religious institutions (15.8%), grocery stores (14.1%), recreational areas (7.2%), healthcare facilities (5.8%), greenspaces (5.6%), and educational institutions (5.5%). In line with Huang et al. (2022), this study has chosen the 12 most populous US cities, including San Francisco, Phoenix, Philadelphia, New York, Miami, Los Angeles, Houston, DC Arlington, Dallas, Chicago, Boston, and Atlanta. The socio-demographic data are sourced from the 2021 American Community Survey (ACS) 5-year estimates and include variables such as median income, and proportion of White, African American, Asian and Hispanic/Latino population.

---

[1] https://www.safegraph.com/

*2.2 Indicators to portray a 15-minute city*

In Figure 1 (a), the black color block means the CBG, while the green color block represents the catchment areas accessible by cycling within a 15-minute radius for residents within this CBG. The grey scatter represents all POI supplies, and the red scatter denotes the POIs visited by residents of this CBG. Moving on to Figure 1 (b), it illustrates a scenario where activities beyond the 15-minute cycling radius can still be accommodated by visiting alternative POIs within the same 15-minute catchment areas. The alternative POIs align with the sub-category of the original activity-related POIs, and should not have been visited by the CBG's residents. The sub-category, in this context, refers to the finest category of POI functions provided by SafeGraph. For example, sub-categories of health activity locations include health facilities, offices of physicians (except mental health specialists), offices of dentists, pharmacies, drug stores, and offices of mental health practitioners (except physicians). Moreover, the number of matched activities should not exceed the capacity of the alternative POI. Since the capacity of POIs is unavailable in this dataset, the total visits to the POI are used as a proxy for the capacity.

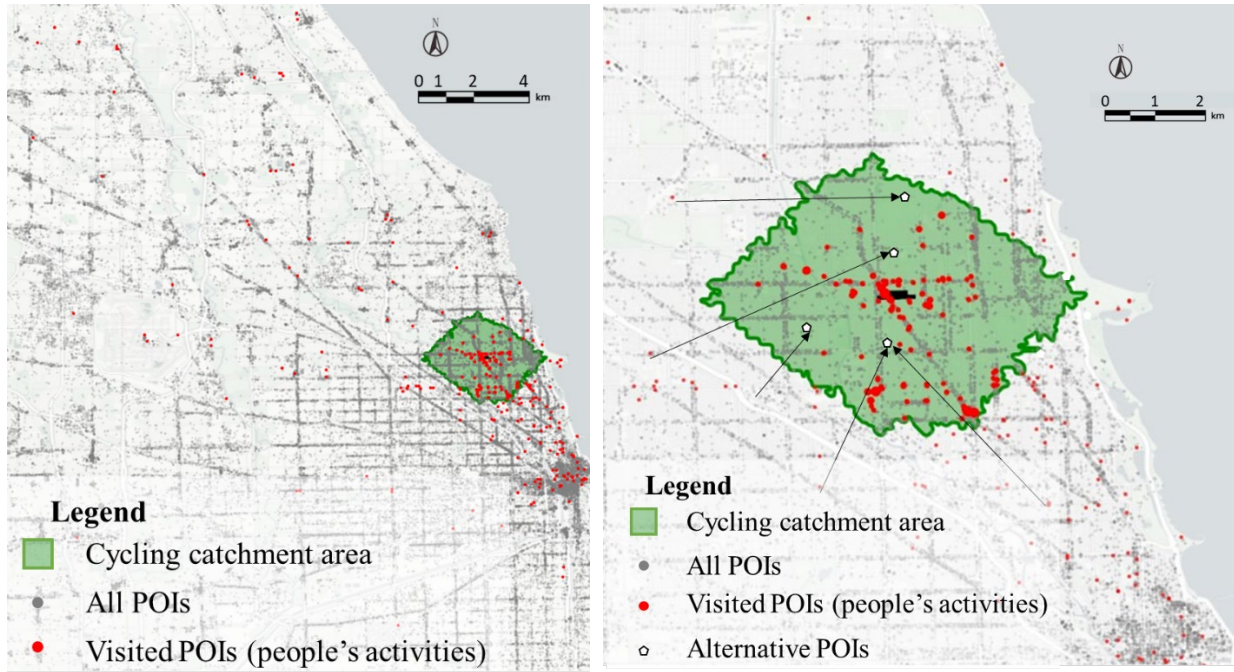

(a) Illustration of accessible POIs and activities conducted within 15 minutes

(b) Illustration of activities beyond the 15-minute cycling that can be satisfied by visiting alternative POIs within 15 minutes

**Figure 1. Illustration of measurement of 15-minute city concept**

Five indicators are used to comprehensively portray a 15-minute city, including *Num_POI* (the number of accessible POIs), *Pct_act_15min* (the percentage of activities conducted within 15 minutes, Equation 1), *Pct_act_sat_15min* (the percentage of activities that can be satisfied within 15 minutes, Equation 2-3), *Pct_reduced_dist* (the proportion of reduced trip distances if people choose to conduct their daily essential activities within 15-minute area, Equation 4), and *Pct_reduced_carbon* (the proportion of reduced $CO_2$ emissions associated with the reduced trip distances, Equation 5). The formulations of these indicators are given as follows, respectively:

$$Pct\_act\_15min = \frac{act_{within\_15min}}{act\_city} \qquad (1)$$

where $act_{within\_15min}$ is the number of activities conducted by residents in a CBG within 15 minutes, $act\_city$ is the total number of activities conducted by residents in a CBG within the city boundary.

$$Pct\_act\_sat\_15min = \frac{act_{within\_15min} + \sum_{j=1}^{J} act_{alternative\_j}}{act\_city} \qquad (2)$$

$$act_{alternative\_j} = \begin{cases} POI_{alternative\_j}, & if\ POI_{alternative\_j} < act_{out\_15min\_j} \\ act_{out\_15min\_j}, & if\ POI_{alternative\_j} \geq act_{out\_15min\_j} \end{cases} \qquad (3)$$

where $j$ is the type of essential functions, including restaurants, service, religious, grocery, recreation, health, greenspace and education. $act_{alternative\_j}$ is the number of activities without 15-minute reach that can be satisfied by finding alternative POIs within 15-minute reach. $POI_{alternative\_j}$ are the capacity of all alternative POIs that are not visited by this home CBG, with the capacity represented by the total visits to these POIs. $act_{out\_15min\_j}$ is the number of activities that are not within 15-minute reach.

$$Pct\_reduced\_dis = \frac{\sum_{j=1}^{J}(\frac{act_{alternative\_j}}{act_{out\_15min\_j}} * dist_{act_{out\_15min\_j}}) - \sum_{j=1}^{J} dist_{POI_{alternative\_j}}}{dist_{city}} \qquad (4)$$

where $dist_{act_{out\_15min\_j}}$ is the total distances between the centroids of CBGs and activity locations that are not within 15-minute reach, $dist_{POI_{alternative\_j}}$ is the total distances between the centroid of CBGs and all alternative POIs, $dist_{city}$ is the total trip distances between current activities of this CBG distances and the centroid of CBG.

$$Pct\_reduced\_carbon = \tag{5}$$

$$\frac{\sum_{j=1}^{J}(\frac{act_{alternative\_j}}{act_{out\_15min\_j}}*dist_{act_{out\_15min\_j}})*Carbon_{car}-\sum_{j=1}^{J}dist_{POI_{alternative\_j}}*Carbon_m}{dist\_act_{out\_15min}*Carbon_{car}+dist\_act_{within\_15min}*Carbon_m}$$

where $Carbon_m$ is the carbon footprint per person per kilometer (i.e., car 197g, public transit 105g (Our World in Data., 2020), walking 26g (Stott, 2020) and cycling 21g (Mizdrak, et al., 2020)). In this case, we calculate the potential reduction in $CO_2$ emission when traveling to an alternative POI within the 15-min reach with respect to different travel modes (i.e., walking, cycling, public transit and car). We assume that activities within 15-minute reach by different travel modes (i.e., walking, cycling, public transit and car) are conducted using this certain travel mode, while activities outside the 15-minute reach are conducted by car. It should also be noted that this indicator may be somewhat underestimated since people may still use cars to access activity locations within 15-minute cycling and public transit distance due to their high dependence on auto vehicles of Americans.

*2.4 Catchment areas by different travel modes*

Previous studies often relied on buffer areas with a Euclidean distance as the travel distance to define the catchment area of a 15-minute city (Di et al., 2023; Ferrer-Ortiz et al., 2022), overlooking the fact that people primarily use navigable road networks for their journeys. Catchment areas used in this research are based on travel time estimation by walking, cycling, public transit, and driving using HERE API[2] (as depicted in Figure 2). HERE API provides more accurate travel time estimation based on comprehensive transportation systems including detailed navigable road networks with precise geometry, extensive link attributes, as well as public transit routes, and transit-related information. The attributes hold substantial significance for analyzing the accessibility of a specific POI to various modes of travel. HERE Isoline Routing API[3] employed in this paper can generate the reachable polygon areas given a certain location (i.e., the centroid of CBG in this paper) and a certain travel mode by setting the travel time parameter (i.e., 15 minutes in this paper).

---

[2] https://developer.here.com/documentation

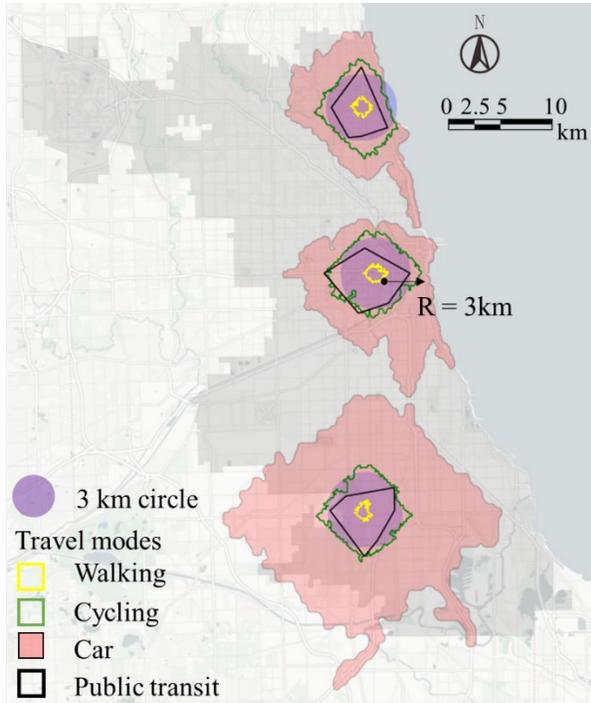

**Figure 2. Illustrations of catchment areas by walking, cycling, public transit and car**

Figure 2 illustrates the catchment areas by multiple travel modes of three different CBGs. The catchment areas for driving differ substantially among various CBGs, primarily due to the consideration of road speeds when estimating the extent people can travel within a 15-minute window. For example, driving speed limit in suburban areas is higher than that in city central areas. As expected, the catchment areas for walking are the smallest. It is worth noting that the catchment areas for cycling tend to be generally larger than those for public transit, suggesting that cycling could be a more efficient mode of transportation in cities examined in this study. The possible reasons include that cycling can offer access to areas that may not be effectively served by public transit. Cyclists have the high flexibility to select their routes, resulting in more direct pathways to their destinations. In addition, public transit systems may not always have stops conveniently located in proximity to all POIs, which leads to considerable walking distances for those relying on public transit to reach their desired destinations.

---
[3] https://developer.here.com/documentation/isoline-routing-api/dev_guide/index.html

*2.3 The Gini coefficient to assess the inequity*

The Gini coefficient is a mathematical metric to represent the overall degree of inequality (Dorfman, 1979). The lower the Gini coefficient, the more equal the distribution of the corresponding variable is across a group of population. Specifically, the Gini coefficient can be calculated as:

$$G = 1 - \sum_{k=1}^{n}(X_k - X_{k-1})(Y_k - Y_{k-1}) \tag{6}$$

where $X_k$ is the cumulated proportion of the population, for $k = 0, ..., n$, with $X_0 = 0$, $X_n = 1$ and $Y_k$ is the cumulated proportion of corresponding variable (i.e., *Num_POI, Pct_act_15min, Pct_act_sat_15min, Pct_ reduced_dist and Pct_reduced_carbon*), for $k = 0, ..., n$, with $Y_0 = 0$, $Y_n = 1$.

## 3. Is the 15-Minute City Concept Within Reach?

By examining the supplies and demands of activities within 15 minutes, as well as the potential to implement a 15-minute city, some interesting findings are reached. First and foremost, residents in the selected cities can only access an average of 47 essential POIs (for brevity, POIs mentioned in the following manuscript refer to essential POIs) within a 15-minute walking distance. This number increases to 449 POIs with 15-minute access via public transit and surges to 846 when considering cycling. A whopping 4028 POIs become reachable within a 15-minute drive. Secondly, only 10% of people's essential activities are within a 15-minute walking distance, whereas more than 65% can be fulfilled within a 15-minute drive. The value is 31.5% for public transit and 41.5% for cycling. The current activity pattern shows that residents in the selected cities are predominantly living in a 15-minute car-centric urban environment, with 65% of their activities catered to within a 15-minute driving distance. Thirdly, when further investigating the potential of implementing a 15-minute city, almost all activities can be satisfied within 15 minutes of driving, public transit, and cycling. Even 15-minute walking can satisfy 84% of residents' essential activity demands. Lastly, embracing the idea of conducting daily essential activities within a 15-minute window can generate substantial environmental benefits. Approximately 70% to 75% of trip distances can be reduced for walking, cycling, and public transportation. In the case of 15-minute driving, about 50% of trip distances can be curtailed. Most notably, the establishment of a 15-minute cycling city emerges as the most potent strategy for reducing $CO_2$ emissions. Given that nearly all essential activities can be satisfied within a 15-minute cycling radius, and cycling is an eco-friendly mode of transportation, this approach can yield an impressive 80% reduction in current $CO_2$ emissions. While some may contend that complete abandonment of motorized travel is unrealistic, our analysis indicates that substantial reductions in $CO_2$ emissions, up to 88% for a 15-minute public transportation city and 50% for a 15-minute car-centric city, can be achieved simply by localizing daily essential activities through the utilization of public transit and private vehicles.

## 4. The Inter-city and Inter-group Comparisons on Assessment of 15-Minute City

*4.1 Supplies of essential POIs within 15 minutes*

Across all the selected cities, people can access the highest number of POIs within 15 minutes by car, followed by cycling. Walking allows access to only a limited number of POIs. When comparing these studies, Los Angeles stands out as the top performer in terms of accessible POIs by car, followed by New York City, San Francisco, and Houston. DC Arlington exhibits the lowest performance in this regard, as shown in Figure 3. In the realm of sustainable transportation alternatives such as cycling, public transit, and walking, people living in New York City can access more POIs compared to other cities. While cities like Atlanta, Dallas, Houston, and Phoenix exhibit the lowest performance on accessible POIs by sustainable travel modes. This is not surprising, given the predominant car-centric nature of these southern US cities.

The calculated Gini coefficients in Figure 3 suggest that the issue of accessibility presents varying degrees of inequality, with the most severe disparities observed in the number of accessible POIs within a 15-minute walking distance. Following this, public transit, cycling, and car accessibility exhibit progressively lesser disparities. Miami emerges as the city with the most equitable distribution of accessible POIs within a 15-minute drive, cycle, and public transit ride. Chicago boasts the most equitable distribution when it comes to 15-minute walking access. Conversely, Atlanta ranks lowest in equitable distribution across the board for 15-minute walking, cycling, and public transit. Although Los Angeles has the highest number of accessible POIs by car, it concurrently faces the most challenging equity issue concerning car accessibility.

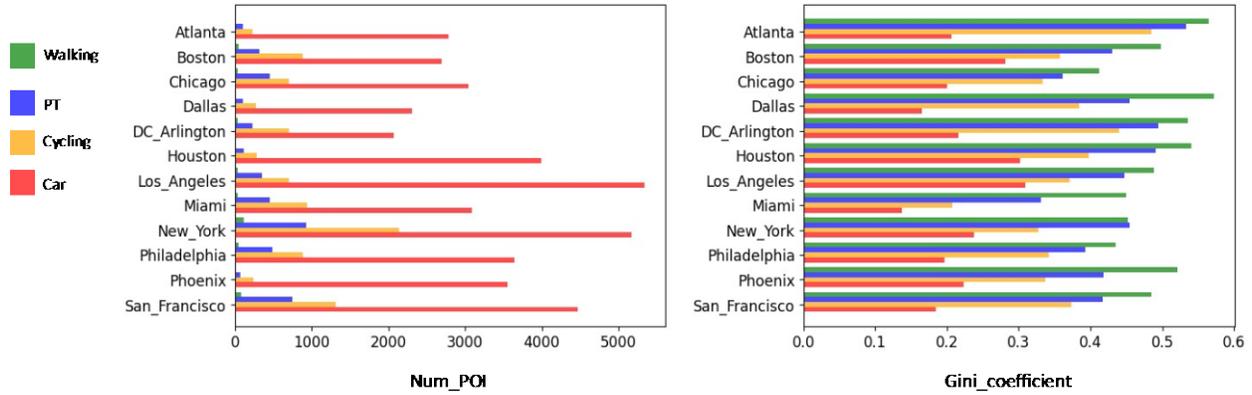

**Figure 3. The number of accessible POIs within 15-minute by multiple travel modes and corresponding Gini coefficient across different cities (PT: public transit).**

Figure 4 presents the bivariate relationships between socio-demographics and accessible POIs by four types of travel modes. Only Pearson's correlation coefficients larger than 0.3 or smaller -0.3 are discussed, with weak correlations are not discussed. It seems that the number of accessible POIs in Los Angeles, New York City, Philadelphia, and Phoenix does not exhibit a strong and moderate correlation with socio-demographics. In most other cities, a positive correlation emerges between the number of accessible POIs within the 15-minute city and the proportion of the White population. The positive correlation is observed in cities like San Francisco, Houston, DC Arlington, Dallas, Chicago, Boston, and Atlanta. A negative correlation is noted with the proportion of African Americans in cities such as Atlanta, Boston, Chicago, Dallas, and DC Arlington. Median income shows a positive correlation with accessible POIs in cities such as Chicago, DC Arlington, and Houston. Notably, an intriguing finding emerges in the case of Miami, where median income and the proportion of the White population exhibit negative correlations with accessible POIs for walking and driving, respectively. This suggests that neighborhoods with higher median income and a greater proportion of the White population are less likely to have a high number of accessible POIs within a 15-minute radius in Miami.

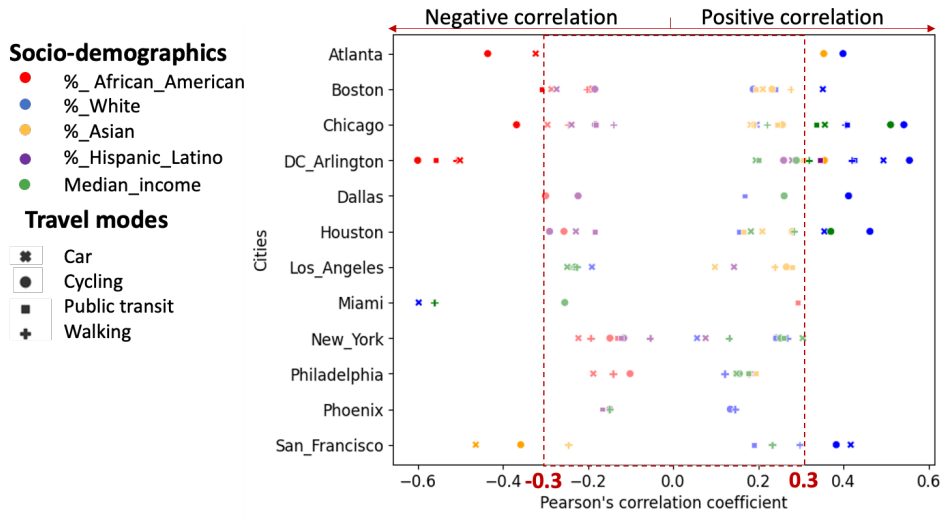

**Figure 4. Pearson's correlation coefficients between the number of accessible POIs and key socio-demographic variables**

*4.2 Activities to essential POIs currently conducted within 15 minutes*

For most selected cities, approximately 60% to 70% of daily essential activities can be reached within a 15-minute drive, a statistic in line with the prevalent reliance on automobiles among Americans. About 30% to over 40% of these activities are accessible via cycling, and roughly 20% to 30% can be reached through public transit. Only around 10% of daily activities fall within walking distance. Remarkably, DC Arlington exhibits the highest proportion of activities conducted within a 15-minute drive, despite having the fewest POIs within this 15-minute range. When considering sustainable travel modes, as shown in Figure 5, DC Arlington, New York City, and Boston stand out with the highest percentages of activities to essential POIs. Cities like Atlanta, Dallas, Houston, and Phoenix demonstrate comparatively lower performance in this regard. New York City shines with the best performance, particularly regarding the percentage of activities attainable within 15 minutes through walking, cycling, and public transit, while Atlanta lags with the least favorable performance. The disparities in activities reachable within a 15-minute drive across various CBGs are relatively less pronounced, with minimal inter-city variations. However, for activities conducted within a 15-minute walk, Atlanta, Dallas, Houston, and Phoenix face the most pronounced inequity challenges.

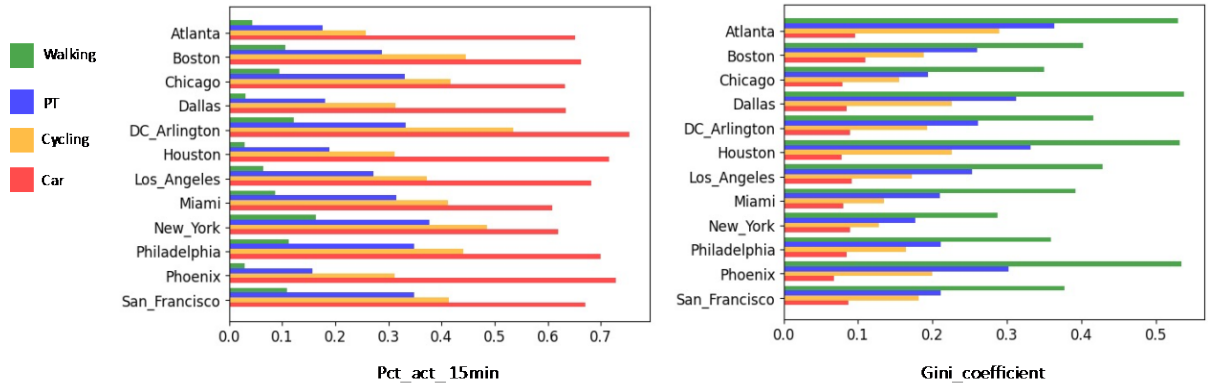

**Figure 5. The percentage of activities within 15-minute by multiple travel modes and the corresponding Gini coefficient across different cities (PT: public transit)**

Figure 6 shows the correlation between the percentage of activities within 15-minute and key socio-demographic variables. Much like the correlation matrix for accessible POIs, the cities of Los Angeles, New York City, and Phoenix do not exhibit a significant or moderate correlation with socio-demographic factors. However, there is an intriguing exception: the proportion of African Americans demonstrates a moderate to even strong negative correlation with the percentage of activities conducted within a 15-minute radius. This suggests that African Americans are less inclined to carry out their daily essential activities within this short timeframe and tend to travel greater distances to access essential services. Except for the proportion of African Americans, all other factors display a positive correlation. This implies that residents in neighborhoods characterized by higher median incomes and a greater proportion of White, Asian,

Hispanic, and Latino populations are more likely to complete their essential daily activities within a 15-minute radius.

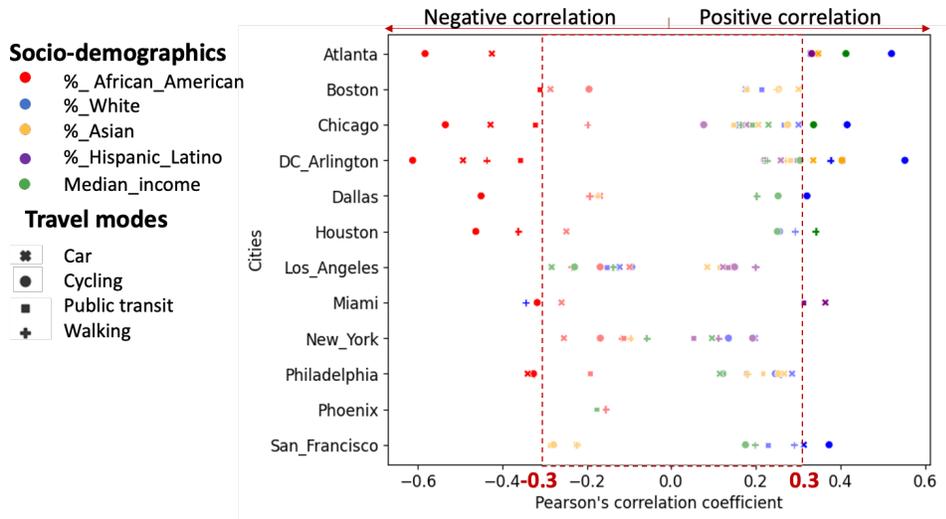

**Figure 6. Pearson's correlation coefficients between the percentage of activities within 15-minute and key socio-demographic variables**

*4.3 The potential of implementing the 15-minute city concept*

When considering the feasibility of introducing the 15-minute city concept in the US, it becomes evident that nearly all essential activities can be conveniently accessed within a 15-minute radius by driving, cycling, or utilizing public transit. As depicted in Figure 7, some cities excel in the concept of the 15-minute walking city, satisfying over 80% of activity requirements, but others lag with satisfaction rates falling below 40%. The extent to which POIs can cater to people's activity needs serves as a crucial indicator of the viability of implementing the 15-minute concept. The development of localized neighborhoods accessible by car, cycling, and public transit has the potential to substantially enhance overall satisfaction levels and serve as an intermediate step toward building a 15-minute walking city.

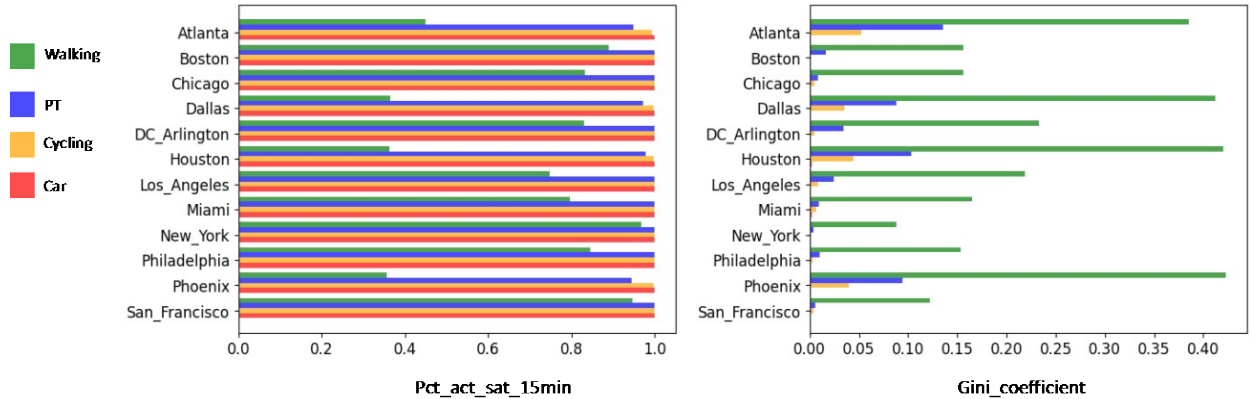

**Figure 7. The percentage of activities that can be satisfied within 15-minute by multiple travel modes and corresponding Gini coefficient across different cities (PT: public transit)**

The 15-minute walking concept is viable for some selected cities, such as New York City, San Francisco, Boston, and Chicago. However, in cities like Atlanta, Dallas, Houston, and Phoenix, the viability of achieving a 15-minute walking city is undermined by the limited percentage of activities that can be accommodated within this short walking radius. While the disparity in the percentage of activities within reach by car versus cycling remains relatively modest between cities, the gap widens significantly when considering public transit. The contrast in the potential for establishing a 15-minute car city is the least pronounced across different cities, whereas the challenge of creating a 15-minute walking city is the most formidable. Turning to the potential for implementing a 15-minute public transit and cycling city, Atlanta emerges as the city facing the most substantial equity disparity, followed by Houston, Dallas, and Phoenix.

When scrutinizing the correlation between socio-demographic factors with the number of accessible POIs (Figure 4) or the proportion of activities completed within a 15-minute radius (Figure 6), several socio-demographic factors reveal varying degrees of correlation, ranging from moderate to strong. However, regarding correlations between socio-demographic factors with the percentage of activities that can be potentially satisfied within 15 minutes (Figure 8), only DC Arlington and San Francisco exhibit moderate correlations. There may be disparities, with African Americans potentially facing challenges and the White population possibly benefiting in terms of POI accessibility and the percentage of activities attainable within 15 minutes, it appears that the potential for implementing a 15-minute city concept does not exhibit significant variations across different socio-demographic groups.

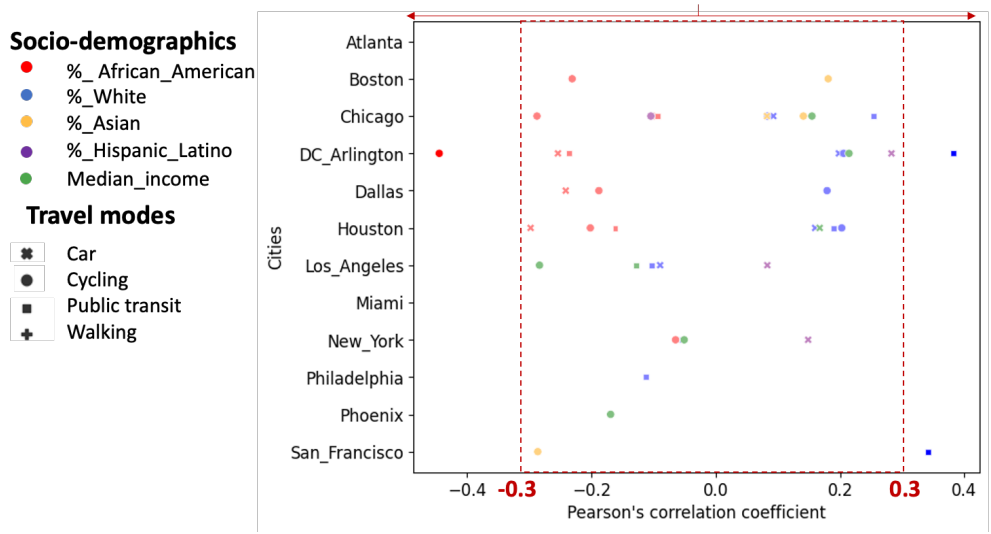

**Figure 8. Pearson's correlation coefficients between the percentage of activities that can be satisfied and socio-demographic variables**

5. **To What Extent Will the 15-Minute City Concept Lead to Reductions in Trip Distance and $CO_2$ Emissions?**

*5.1 Estimated trip distance reduction*

Promoting the concept of a 15-minute city holds substantial potential for reducing both travel distances and $CO_2$ emissions. Both 15-minute cycling and public transit city share similar potential to reduce more trip distances compared to the 15-minute city model centered around cars and walking. The promotion of localized neighborhoods accessible by car may result in more POIs becoming reachable within a 15-minute driving radius, and a larger share of activities conducted within. Although it is less sustainable than other modes, it still has the potential to reduce about 30% to 50% of trip distances. In addition, Figure 9 shows that apart from cities that are not suitable for implementing 15-minute walking cities, such as Atlanta, Dallas, Houston, and Phoenix, adopting the 15-minute walking city can reduce more trip distances compared to a 15-minute car city. In contrast, the issues related to inequity in travel distance reductions are less pronounced for 15-minute public transit and cycling cities compared to their car-centric counterparts.

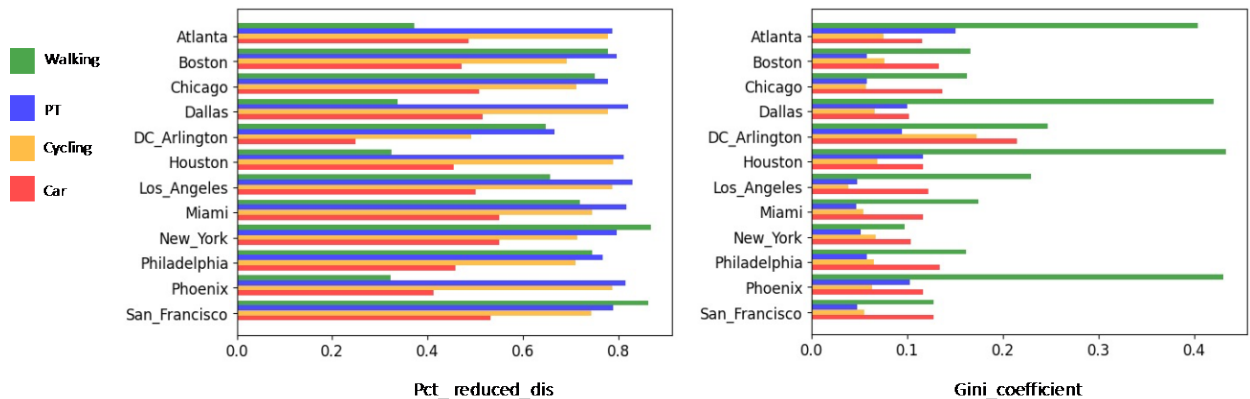

**Figure 9 The percentage of trip distance reduction within 15-minute by multiple travel modes and corresponding Gini coefficient across different cities (PT: public transit)**

In the majority of cities, such as Atlanta, Houston, Los Angeles, New York City, Philadelphia, and Phoenix, the percentage of trip distance reduction does not exhibit significant differences across various socio-demographics, as illustrated in Figure 10. For Boston, Chicago, and Miami, median income and the proportion of the White population show notable correlation with the percentage of trip distance reduction. However, in Dallas, these two factors display a negative correlation with trip distance reduction. This suggests that in Boston, Chicago, and Miami, neighborhoods with higher median income and a greater proportion of the White population are more likely to contribute to a more substantial reduction in trip distances, whereas in Dallas, these neighborhoods are less likely to contribute to higher trip distance reduction. In the context of a 15-minute cycling city, African Americans in DC Arlington are inclined to contribute to greater trip distance reduction, but this propensity decreases when transitioning to a 15-minute public transit city. Moreover, the White population in DC Arlington is more likely to reduce trip distances in a 15-minute public transit city, but their contribution diminishes when transitioning to a 15-minute cycling city.

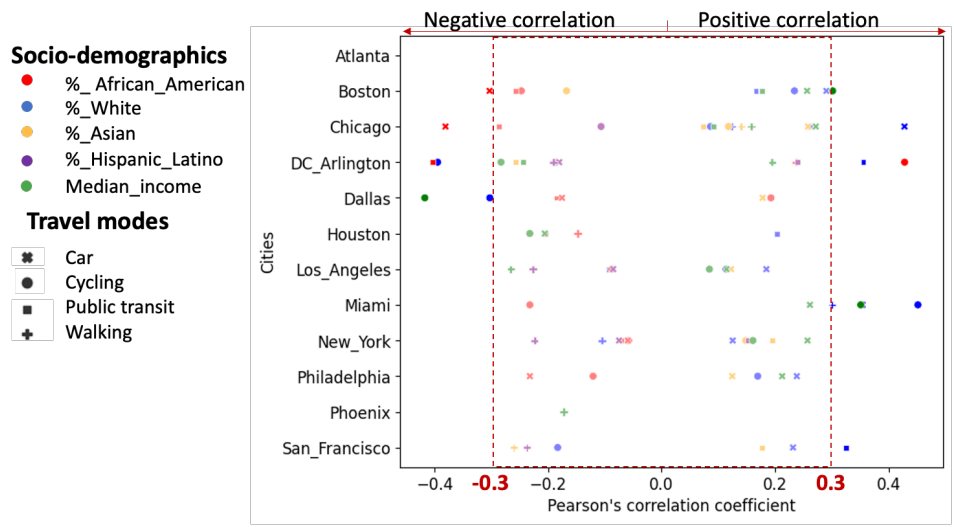

**Figure 10 Pearson's correlation coefficients between the percentage of trip distance reduction and socio-demographic variables**

*5.2 Estimated $CO_2$ emission reduction*

In most urban settings, opting for a 15-minute public transit or cycling city proves to be a more effective strategy for reducing $CO_2$ emissions compared to walking and driving, with the notable exceptions of New York City and San Francisco. As shown in Figure 11, in the aforementioned two cities a 15-minute walking city outperforms public transit regarding the potential of $CO_2$ emission reduction. However, in cities like Atlanta, Dallas, Houston, and Phoenix, relying on a 15-minute walking city is less promising since it yields the smallest reduction among all transportation modes. The disparity in $CO_2$ emission reductions when transitioning to a 15-minute walking city is more pronounced than for other modes of travel, with Atlanta, Dallas, Houston, and Phoenix experiencing the most pronounced disparities. The equity concerns associated with a 15-minute cycling city are less severe when compared to other transportation modes, suggesting that cycling presents a relatively equitable approach to reducing emissions across a variety of urban contexts.

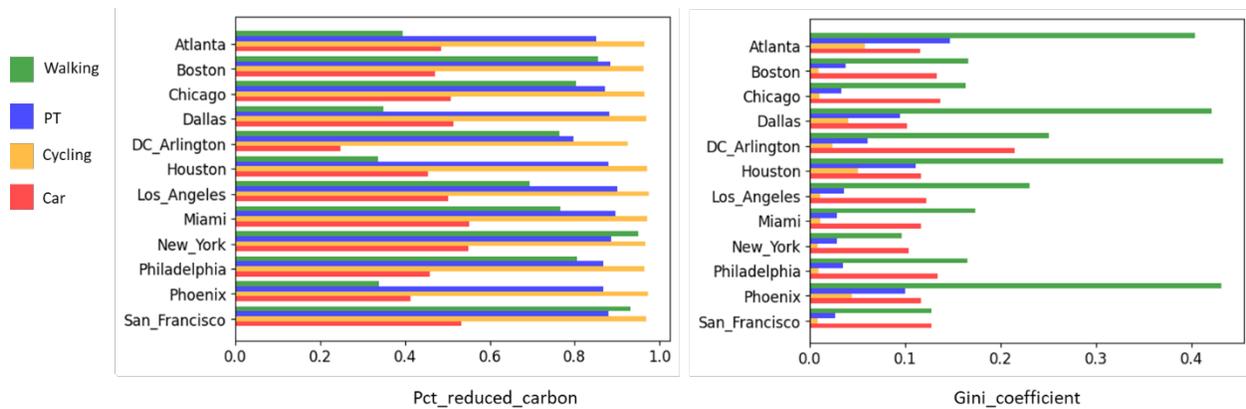

**Figure 11 The percentage of $CO_2$ emission reduction within 15-minute by multiple travel modes and corresponding Gini coefficient across different cities (PT: public transit)**

Interesting findings can be reached when comparing the correlation matrix of potential trip distance reduction and $CO_2$ emission reduction. While mixed findings are reached regarding the correlation between reduced trip distances and socio-demographics, the proportion of the White population is positively correlated, and the proportion of African Americans is negatively correlated with potential $CO_2$ emission reductions (Figure 12). For Atlanta, Dallas, Los Angeles, New York City, Philadelphia, and Phoenix, it appears that the percentage of $CO_2$ emission reduction remains relatively consistent across various socio-demographic groups, suggesting that this potential reduction is not strongly influenced by demographic factors in these cities.

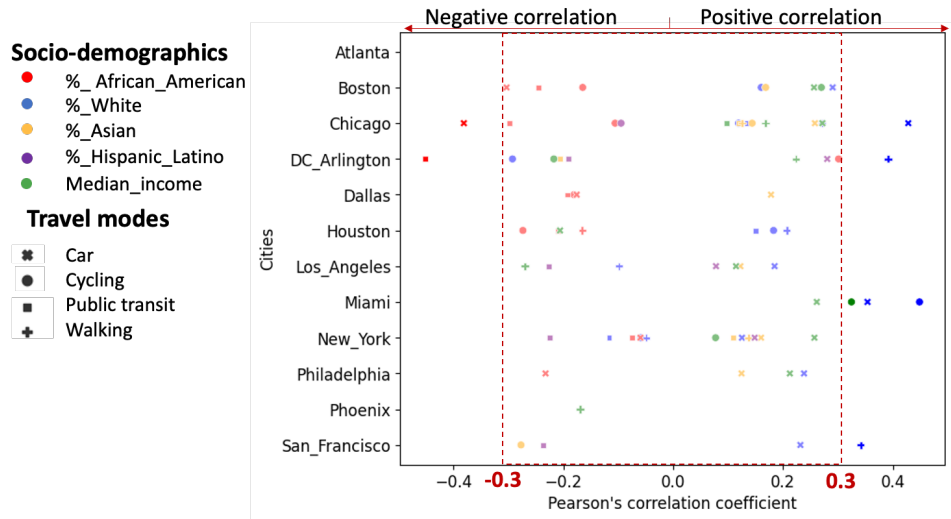

**Figure 12 Pearson's correlation coefficients between the percentage of $CO_2$ emission reduction and socio-demographic variables**

## 6. Conclusions

The realization of self-sufficient neighborhoods, as envisaged by the 15-minute city concept, holds the promise of reducing travel distances and shifting away from motorized transportation in favor of active modes of travel. This shift can, in turn, contribute significantly to the reduction of $CO_2$ emissions. Empirical studies have underscored that households residing in walkable neighborhoods tend to experience significantly lower transportation costs when compared to those in car-dependent areas. Evaluating the combined potential reduction in trip distances and $CO_2$ emissions resulting from the 15-minute city concept is of great value, offering profound insights into the potential benefits of fostering localized communities (Li et al., 2023).

Recognizing that the complete elimination of vehicle-based mobility may not be feasible, the assessment of the 15-minute city concept encompasses both active modes of travel (such as walking and cycling) and motorized modes (including public transportation and driving). This comprehensive approach allows for a thorough evaluation of the concept's impact on transportation inequality and environmental sustainability.

In selected cities, the prevalence of car dependency arises from the fact that most daily activities extend beyond a 15-minute walking, cycling, or public transit radius. However, with the implementation of the 15-minute city concept, a majority of these activities can indeed be accommodated within a 15-minute reach, whether by cycling, using public transportation, or driving. Encouraging residents to conduct their essential activities through cycling and public transit emerges as the most promising means of reducing trip distances and $CO_2$ emissions for most cities studied in this paper. The 15-minute walking concept is also feasible in a few cities, such as New York City, San Francisco, Boston, and Chicago. Due to the low percentage of activities that can be satisfied within a 15-minute walking distance, Atlanta, Dallas, Houston, and Phoenix are not within reach of the 15-minute walking city. Given the fact that Americans heavily rely on car usage, promoting localized 15-minute driving neighborhoods remains meaningful, potentially leading to an efficient potential reduction of approximately 50% of trip distances and $CO_2$ emissions.

On the one hand, inter-group comparisons reveal disparities, with neighborhoods having a higher proportion of White population and higher median income enjoying greater accessibility to POIs and a higher percentage of activities within the 15-minute radius. On the other hand, African Americans experience fewer accessible POIs and a lower percentage of activities. As a result, when implementing the 15-minute city concept given the current service supply, neighborhoods with a higher percentage of White population may have a higher potential to contribute to $CO_2$ emissions reduction compared to neighborhoods with more African Americans.

Admittedly, implementing the 15-minute city concept in US cities is not without its challenges (Khavarian-Garmsir et al., 2023). Some neighborhoods may lack the necessary walkable and bike-friendly infrastructure, which is not accounted for in this study. Adapting existing cities to the ideals of the 15-minute city, especially for walking and cycling, may entail substantial transformations and changes to the urban landscape. Moreover, the concept's applicability varies based on the unique geographic, social, and economic contexts of different cities, necessitating a tailored approach for each community and neighborhood. Furthermore, this research focuses primarily on US megacities, and comparisons among cities of different scales may provide valuable insights, a subject worthy of future research. The 15-minute city concept is also somewhat generic and can be adapted to different timeframes, such as 20 or even 30 minutes (Logan et al., 2022).

Despite the above limitations, this study comprehensively evaluated the 15-minute city concept in the 12 most populous cities. The findings gained from this research can offer policymakers and urban planners an insight into how far US cities are away from the 15-minute city and the potential emission reduction they can expect if the concept is successfully implemented.

Arizona State University. (2021). Tempe: 20-minute city. Transportation. https://transportation.asu.edu/projects/tempe-20-minute-city/.

Büttner, B., & Seisenberger, S., Larriva, M., Gante, A., Ramírez, A., & Haxhija, S. (2022). Urban mobility next 9 ±15-minute city: Human-centred planning in action Mobility for more liveable urban spaces EIT Urban Mobility Munich November 2022. 10.13140/RG.2.2.10482.79041.

City of Portland. (2010). 20-Minute neighborhoods. https://www.portlandonline.com/portlandplan/index.cfm?c=52256&a=288098.

Coleman, N., Gao, X., DeLeon, J., & Mostafavi, A. (2022). Human activity and mobility data reveal disparities in exposure risk reduction indicators among socially vulnerable populations during COVID-19 for five US metropolitan cities. *Scientific Reports*, 12(1), 15814.

Coston, A., Guha, N., Ouyang, D., Lu, L., Chouldechova, A., & Ho, D. E. (2021). Leveraging administrative data for bias audits: Assessing disparate coverage with mobility data for COVID-19 policy. In *Proceedings of the 2021 ACM Conference on Fairness, Accountability, and Transparency* (pp. 173-184).

Di Marino, M., Tomaz, E., Henriques, C., & Chavoshi, S. H. (2023). The 15-minute city concept and new working spaces: A planning perspective from Oslo and Lisbon. *European Planning Studies*, 31(3), 598-620.

Dorfman, R. (1979). A formula for the Gini coefficient. *The Review of Economics and Statistics*, 146-149.
28